\documentclass[journal]{IEEEtran}

\usepackage{amsmath}
\usepackage[switch,pagewise,columnwise]{lineno}

\usepackage{graphicx}

\begin{document}

\title{CCAT: Flexible Stripline Circuits for Large-Format Kinetic Inductance Detector (KID) Array Readout}

\author{Ben Keller, Rodrigo Freundt, James R. Burgoyne, Scott Chapman, Steve Choi, Cody J. Duell,  Christopher Groppi, Caleb~Humphreys, Lawrence T. Lin, Alicia Middleton, Michael D. Niemack, Darshan Patel, Eve Vavagiakis, Samantha Walker, Yuhan Wang, Ruixuan (Matt) Xie

\thanks{B. Keller, C. J. Duell, R. Freundt, C. Humphreys, L. T. Lin, M. D. Niemack, D. Patel, S. Walker, and Y. Wang are with the Department of Physics, Cornell University, Ithaca, NY 14853, USA. (e-mail: bdk54@cornell.edu).

J. R. Burgoyne and R. Xie are with the Department of Physics and Astronomy, University of British Columbia, Vancouver, BC V6T 1Z1, Canada

S. Chapman is with the Department of Physics and Atmospheric Science, Dalhousie University, Halifax, NS B3H 4R2, Canada.

S. Choi is with the Department of Physics and Astronomy, University of California, Riverside, CA 92521, USA. 

C. Groppi is with the School of Earth and Space Exploration, Arizona State University, Tempe, AZ 85287, USA.

E. Vavagiakis is with the Department of Physics, Duke University, Durham, NC, 27704, USA, and the Department of Physics, Cornell University, Ithaca, NY 14853, USA.
}}

\markboth{Journal of \LaTeX\ Class Files,~Vol.~14, No.~8, August~2015}%
{Shell \MakeLowercase{\textit{et al.}}: Bare Demo of IEEEtran.cls for IEEE Journals}

\maketitle

\begin{abstract}

The CCAT Observatory's primary science instrument, Prime-Cam, is nearing readiness for deployment to the Fred Young Submillimeter Telescope (FYST) in the Atacama Desert in northern Chile. When fully deployed, Prime-Cam will field approximately 100,000 kinetic inductance detectors (KIDs) across seven instrument modules making both broadband and polarimetric measurements. Meanwhile, in-lab characterization of the first CCAT instrument module, a 280 GHz broadband camera fielding over 10,000 KIDs, is currently underway in the testbed instrument Mod-Cam. Both Mod-Cam and Prime-Cam will employ 46~cm long low-thermal-conductivity flexible circuits (``stripline") between 4~K and 300~K to connect large-format arrays of multiplexed KIDs in each instrument module to readout electronics. The 280~GHz camera currently installed in Mod-Cam uses six of these striplines to read out its over 10,000 detectors. We have examined the thermal and electrical performance of the stripline installed in Mod-Cam in an effort to better optimize the readout chain for deployment in Prime-Cam. We begin by characterizing the OFHC copper in the stripline traces, allowing for the estimation of thermal loading through these flexible circuits in their configurations in both Mod-Cam and Prime-Cam. We then directly measure the thermal conductivity of the stripline, finding it is best described by $kA~=~22\pm6~T^{0.84\pm0.09}~\mathrm{\mu W~m~K^{-1}}$ for temperature ranges of $6~\mathrm{K} < T <  20~\mathrm{K}$ and $kA~=~0.6\pm0.3~T^{-0.4\pm0.1}~\mathrm{mW~m~K^{-1}}$ for ranges from $20~\mathrm{K}< T < 80~\mathrm{K}$. Following our thermal characterizations, we report on the transmission and crosstalk properties of the Mod-Cam readout chain, isolating elevated crosstalk to SMP-SMA transition printed circuit boards (PCBs) that interface with the stripline. This finding validates the stripline circuit as a viable high-density cabling option for large-format array readout. %Finally, we outline future improvements to the transition PCBs to enhance isolation between adjacent readout chains in both Mod-Cam and Prime-Cam. 

\end{abstract}

\begin{IEEEkeywords}
CMB, KID, readout, flexible circuits, instrumentation
\end{IEEEkeywords}

\IEEEpeerreviewmaketitle

\section{Introduction}\label{section:intro}

\begin{figure}%[!h]
\includegraphics[width=\linewidth]{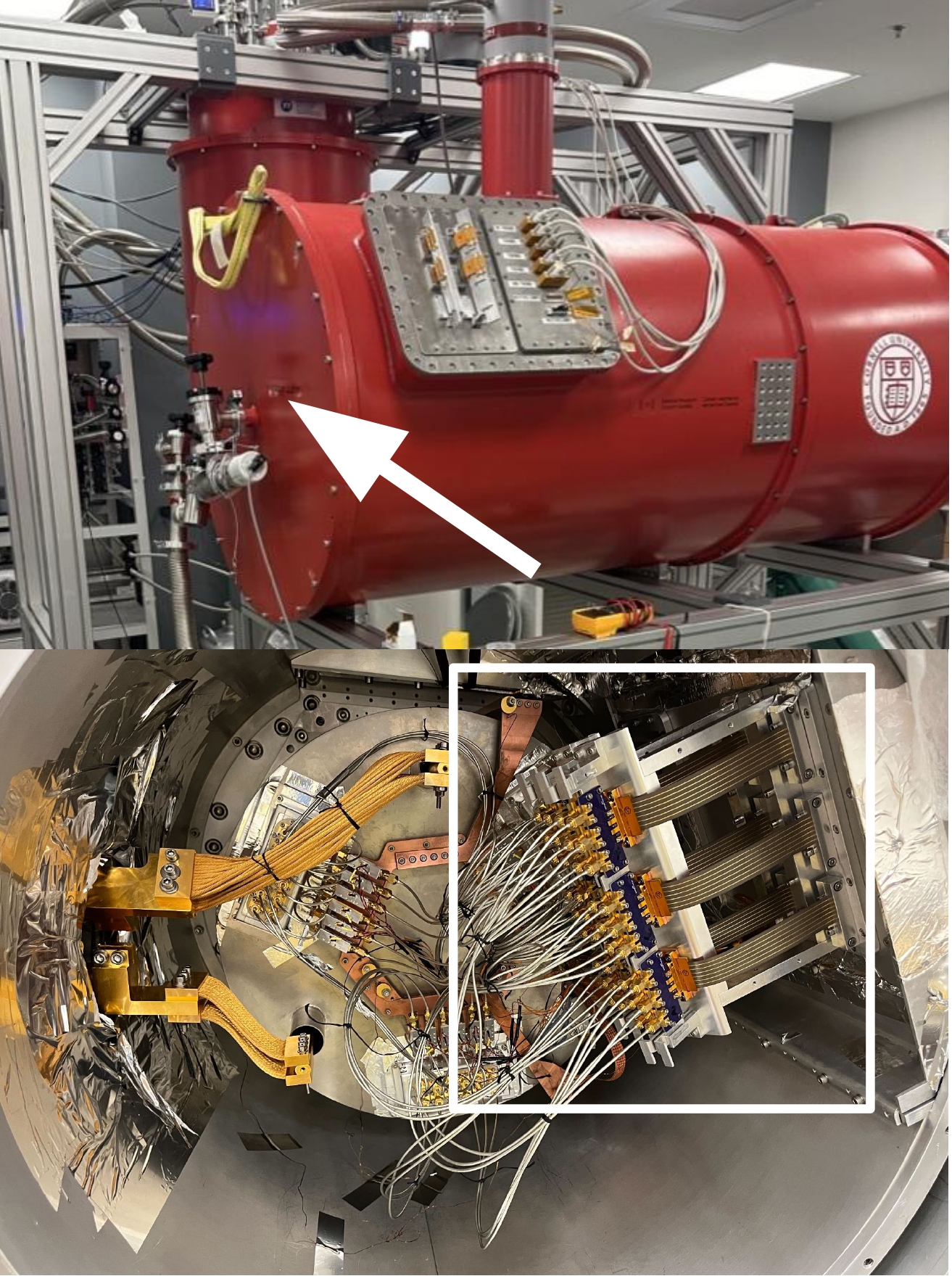}
\includegraphics[width=\linewidth]{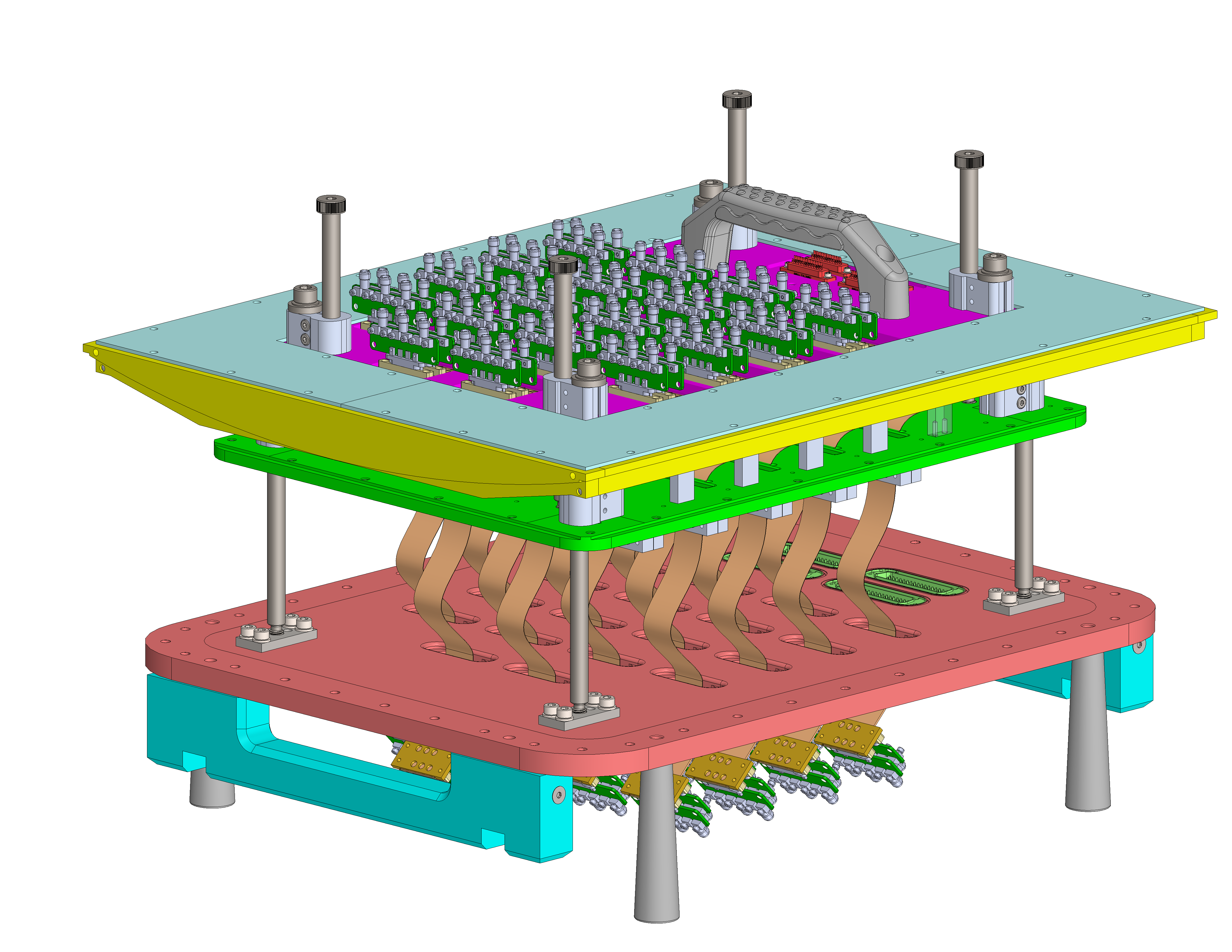}
\caption{\textit{Top}: A picture of Mod-Cam from the side. The white arrow denotes the ``back" of Mod-Cam. \textit{Middle}: The back of Mod-Cam when the cryostat is open, after the instrument module has been installed. The mount for the six stripline assemblies is highlighted in the white box, along with the purple SMP-SMA transition PCBs and coaxial cable interface to the 4~K low noise amplifiers (LNAs). \textit{Bottom}: A preliminary rendering of the larger Prime-Cam readout harness with 18 stripline assemblies and accompanying transition PCBs. The red, green, and yellow plates in the rendering are mounted to the Prime-Cam 300, 40, and 4~K shells respectively. Temporary installation rods shown between plates in the rendering are removed after the readout harness is inserted in Prime-Cam.}
\label{fig:modcam}
\end{figure}

\IEEEPARstart{T}{he} CCAT Collaboration is currently constructing the Fred Young Submillimeter Telescope (FYST), a six-meter off-axis crossed-Dragone telescope, at an elevation of 5600~m on the summit of Cerro Chajnantor in the northern Chilean Atacama Desert \cite{Huber:huber_optics}. Due to the low precipitable water vapor (PWV) at the telescope site \cite{2020_chajnantor_pwv}, FYST will offer an unprecedented view into the sub-millimeter sky, with observations scheduled to begin in 2026. The CCAT Collaboration has been assembling and testing two instruments, either of which may be housed in FYST: the initial testbed instrument Mod-Cam, and the first generation science instrument Prime-Cam. Currently, the CCAT collaboration plans to deploy Prime-Cam to FYST in mid-2026, while Mod-Cam will continue to be utilized for detector and readout testing in the lab. While Prime-Cam assembly and validation is ongoing, Mod-Cam has allowed for the extensive testing of readout hardware and software which will be implemented similarly in Prime-Cam. When deployed, Prime-Cam will be capable of making broadband polarimeteric observations at frequencies ranging between 280-850~GHz with additional spectroscopic surveys performed at 210-420~GHz with $R \sim 100$ \cite{nikola_eorspec}, targeting science goals ranging from understanding the growth of structure during the Epoch of Reionization to constraining cosmology using measurements of the Sunyaev-Zel'dovich (SZ) effect \cite{CCATsciencepaper_2023}.

\begin{figure}%[!h]
\includegraphics[width=\linewidth]{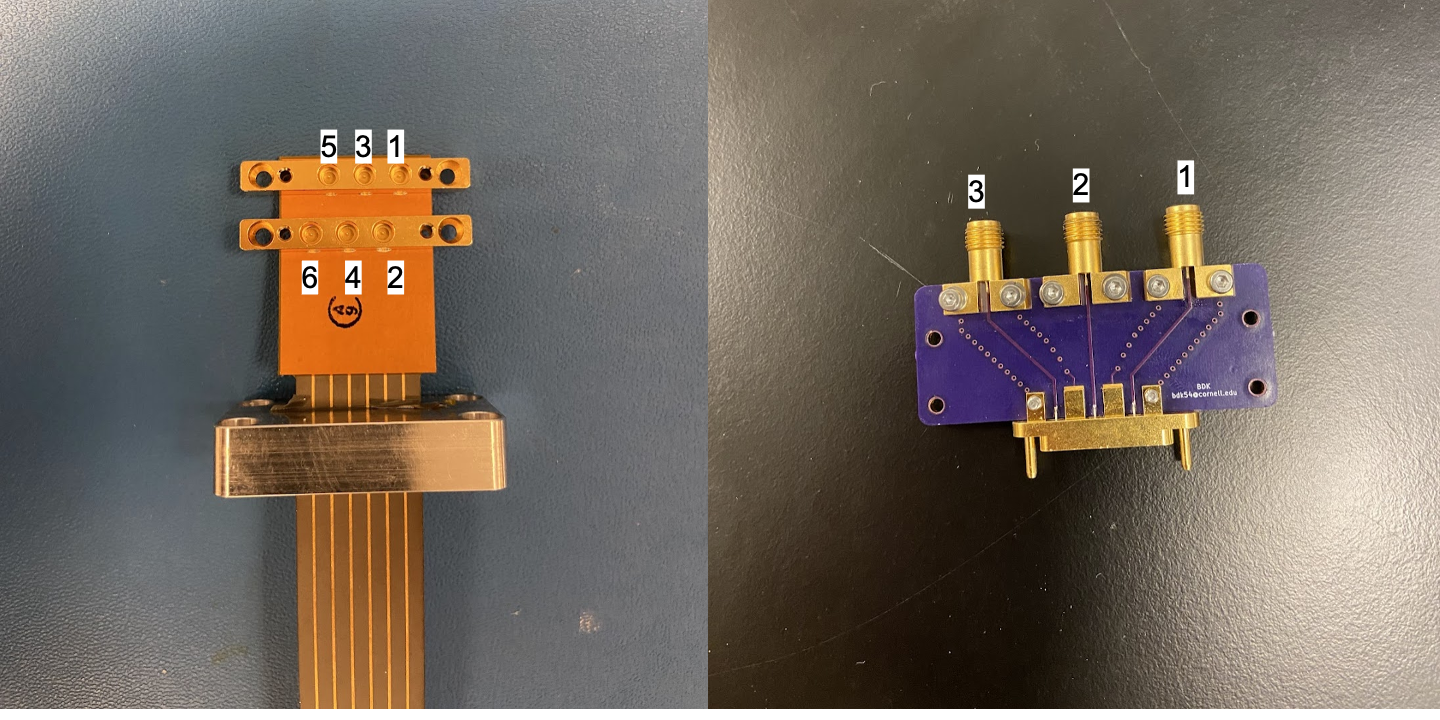}
\includegraphics[width=\linewidth]{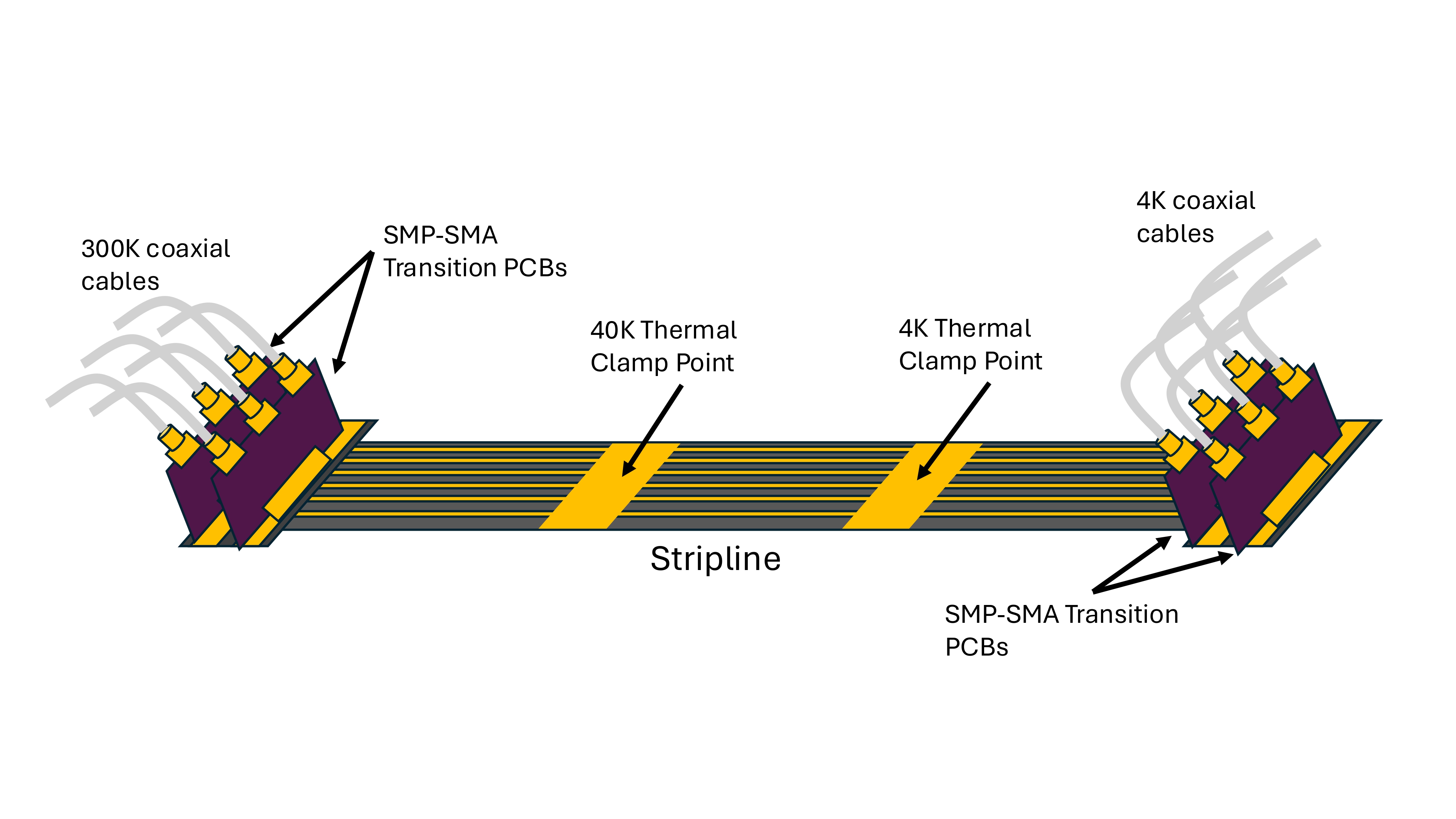}
\caption{\textit{Top Left}: The connectorized end of one stripline circuit. The numbers denote the labeling scheme used between channels for the ``stripline-only" and ``stripline+PCB" crosstalk measurements shown in Fig. \ref{fig:crosstalk}. \textit{Top Right}: A photograph of one SMP-SMA transition PCB.  Two PCBs plug into ganged triple-SMP connectors on each end of the stripline. The numbers overlaid denote the labeling scheme used in the ``PCB-only" crosstalk measurement shown in Fig. \ref{fig:crosstalk}. \textit{Bottom}: A diagram demonstrating the full stripline and SMP-SMA transition used in Mod-Cam. Four identical transition PCBs are used to connect to coaxial cables at 300~K and 4~K.  Thermal clamping points where the ground plane is extended to improve heatsinking are shown on the stripline; these physically coincide with the location of the 40~K and 4~K shells.}
\label{fig:pcb_picture}
\end{figure}

Mod-Cam is a cryogenic receiver with a side-car dilution refrigerator (DR) design featuring space for a single instrument module \cite{Vavagiakis_modcam}. Mod-Cam has been used extensively as an in-lab testbed for the first CCAT instrument module, a 280~GHz broadband camera. The 280~GHz module features over 10,000 KID detectors split across three arrays, each with six networks of $\sim 570$ detectors \cite{Duell_280module}. To connect the detectors to the warm RFSoC readout electronics, Mod-Cam employs six 46 cm low-thermal-conductivity flexible circuits (``stripline") between 4~K - 300~K. At either end of the stripline, SMP-SMA transition printed circuit boards (PCBs) connect ganged triple SMP connectors on the stripline to flexible coaxial cables. These transition PCBs are seen in Fig. \ref{fig:modcam}; Fig. \ref{fig:pcb_picture} also shows a schematic diagram of the full stripline 4~K~-~300~K transition. 

Prime-Cam, CCAT's primary science instrument, has recently been assembled for the first time at Cornell and is currently undergoing cryogenic validation. Prime-Cam features space for seven modules, all with the same mechanical form as the 280~GHz camera. Additional modules currently under construction are the 350~GHz and 850~GHz broadband cameras, as well as the EoR-Spec spectrometer module \cite{huber_850module}\cite{nikola_eorspec}. Additional spectrometer-on-a-chip module designs are planned, bringing the total number of detectors housed in Prime-Cam up to $\sim 100,000$ total KIDs \cite{CCATsciencepaper_2023}. 

Reading out detector counts of this magnitude presents a nontrivial challenge in terms of hardware, both cryogenically and mechanically. Resonator spacing requirements to minimize collisions between adjacent-frequency resonators create an upper bound on the number of detectors read out on a single RF network, while instrument thermal budgets present a constraint on the total number of RF networks that can reasonably be installed without constituting an unacceptably large thermal load on the cold temperature stages. The first three CCAT instrument module designs (280~GHz, 350~GHz, and the EoR-spec spectrometer) baseline $\sim500$~MHz readout bandwidth with $\sim570$ per RF network \cite{sinclair2022}. To read out all detectors from just these first three instrument modules would require 48 RF networks and a total of 192 pieces of coaxial cable. Future instrument modules, such as the 850~GHz broadband module, feature ballooning detector counts and subsequently also the need for many additional RF networks per module. The total quantity of cryogenic coaxial cabling needed in Prime-Cam would push the limits of the instrument thermal budget and physical space constraints, especially at the 4~K stage. With CCAT and other experiments pushing on-sky detector counts higher with large-format KID arrays, flexible stripline and other high-density readout cabling solutions may present and attractive alternative that avoids the pitfalls of standard coaxial cables.

As such, Prime-Cam will employ similar stripline to that installed in the testbed instrument Mod-Cam to read out signals between 4~K~-~300~K. Current plans feature the 280~GHz, 350~GHz, and EoR-Spec modules being read out through one of three available readout ports, or ``readout harnesses", using 18 total striplines. Of these, the three cameras will actively use 48 of the available 54 readout chains (6 chains, or 2 full stripline being reserved as spares). A preliminary design for this first readout harness, along with the SMP-SMA transition PCBs, is shown in the bottom panel of Fig. \ref{fig:modcam}. For the new stripline installed in the Prime-Cam readout harness, changes to the original Mod-Cam circuit geometry have been made to reduce thermal loading at the 4~K stage where the heat budget is limited. The updated designs for the Prime-Cam stripline feature half the copper pour thickness in the signal and ground traces ($9~\mu \mathrm{m}$) along with a roughly doubled length of circuit (20~cm) between the 4~K~-~40~K stages as compared to the designs in Mod-Cam. Fabrication and testing of the Prime-Cam stripline and first readout harness with these modifications is currently underway.

%moved previous fig 5 up here since it is now referenced earlier 

In this work we present an overview of the stripline currently installed in Mod-Cam. We characterize the thermal conductivity of the circuit by measuring both the residual resistance ratio (\textit{RRR}) and the thermal conductivity-temperature relation for the flexible circuits. Using the \textit{RRR} measurement, we obtain thermal loading estimates for the stripline currently installed in Mod-Cam, as well as predictions for the newly modified designs for Prime-Cam. We then present measurements of the room-temperature transmission and crosstalk through the stripline and SMP-SMA transition PCBs used to interface with readout coaxial cables in Mod-Cam, isolating the source of elevated crosstalk in the Mod-Cam readout chains to the transition PCBs.

\section{Mod-Cam Cryogenic Readout and Stripline Design}\label{section:stripline_design}

The 280~GHz broadband instrument module currently installed in Mod-Cam features three KID arrays read out across 18 total RF chains; these RF chains use 304 Stainless Steel coaxial cables on the inputs from 0.1-4~K and NbTi coaxial cables on the outputs from 0.1-4~K \cite{Duell_280module}. There is a mixture of attenuation installed at 0.1~K, 1~K, and 4~K on the input side of the readout chains. Optimization of the chain attenuation for thermal stability of the focal plane and readout noise is ongoing. Cryogenic low noise amplifiers manufactured at Arizona State University are connected to the outputs and physically mounted at 4~K on the back of the instrument module, as seen in Fig. \ref{fig:modcam}. Flexible coaxial cables connect the chains to the stripline installed in Mod-Cam via SMP-SMA transition PCBs shown in Fig. \ref{fig:pcb_picture}. 

\begin{figure}[!]
\includegraphics[width=\linewidth]{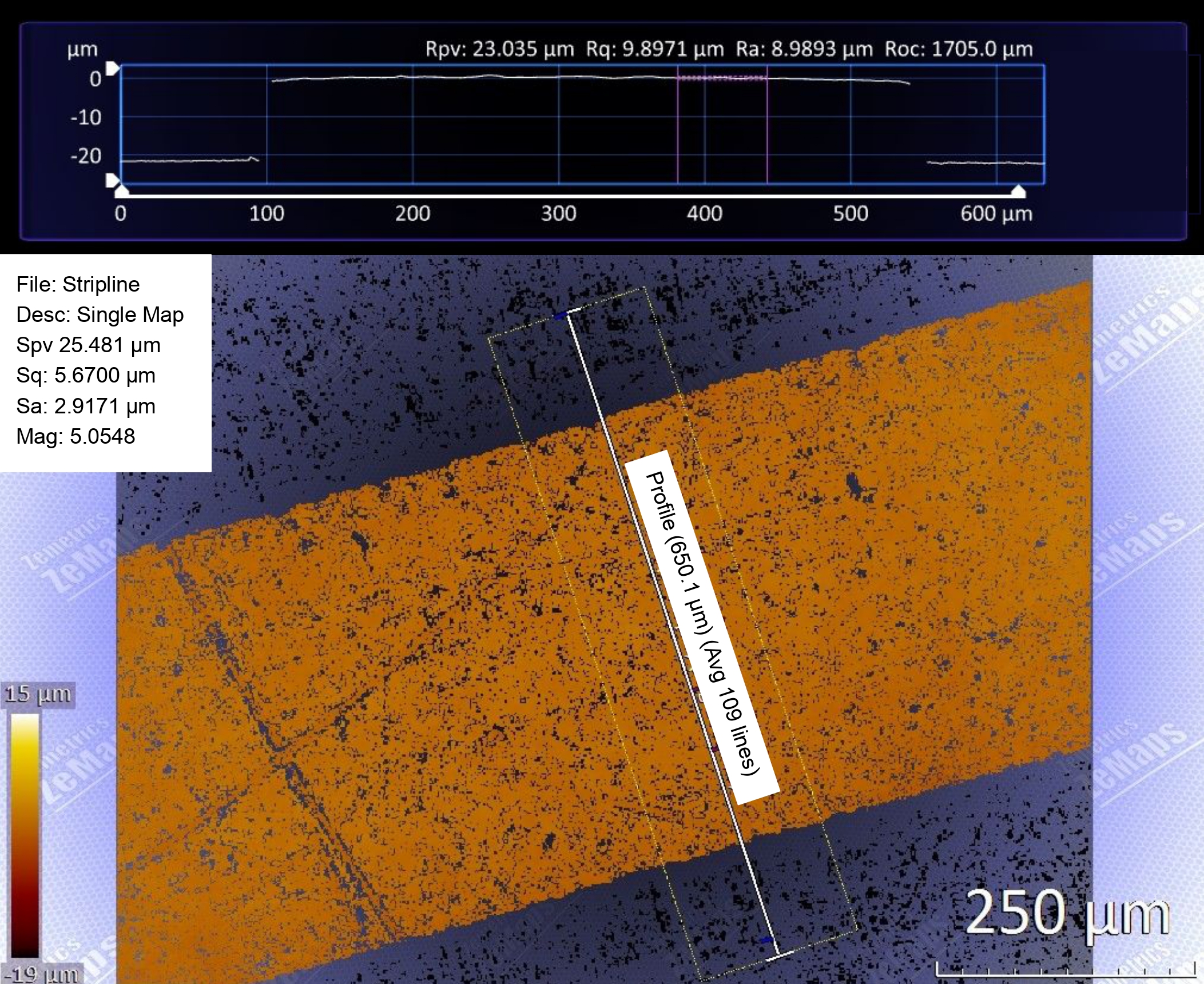}
\caption{Profilometry of a ground trace on a stripline assembly showing the trace thickness to be approximately $22~\mu \mathrm{m}$ thick and $0.43~\mathrm{mm}$ wide.}
\label{fig:profilometry}
\end{figure}

From 4~K to 300~K, Mod-Cam employs six stripline assemblies to bring signals into and out of the instrument module. The stripline characterized in this paper and installed in Mod-Cam are 46~cm long DuPont flexible circuits with a buried copper signal layer sandwiched between polyimide, and outer copper ground layers on either side spanning the length of the circuit \cite{neric_stripline}. At designated clamping points, schematically shown in Fig. \ref{fig:pcb_picture}, the ground plane is extended fully across the width of the circuit to provide better thermal contact with heat sinking clamps at the 4~K and 40~K stages in the cryostat. The ground traces are otherwise kept to $0.43~\mathrm{mm}$ width to minimize thermal conductivity along the length of the circuit. Profilometry in Fig. \ref{fig:profilometry} shows the ground traces are measured to be approximately $22~\mu$m thick. Original designs called for trace dimensions of $18 \mu m$ thickness and $0.38~\mathrm{mm}$ width. The additional cross-sectional area is believed to result from the gold plating being thicker than expected.

The complete stripline assemblies are built at Arizona State University, where each end of the circuit is soldered to two custom Delta RF ganged triple-SMP connectors. The circuit is then epoxied into a vacuum flange at a fixed location. On the back of each connectorized region, a thin G10 tab is added for mechanical rigidity and and to protect the circuit when SMP connections are made and unmade. The stripline interface with coaxial cables at 4~K and 300~K uses a custom transition PCB designed at Cornell that takes the ganged triple-SMP connectors and converts them to SMA connections as shown in detail in Fig.~\ref{fig:pcb_picture}. 

\section{Thermal Characterization of Stripline}
\subsection{RRR Measurement}\label{section:RRR_result}

To measure the Residual Resistance Ratio (\textit{RRR}) value of the copper in the flexible circuit, we cut a section of stripline approximately 5~cm long and installed it on the 4~K stage of a Bluefors LD400 dilution refrigerator. Using a Lakeshore 372 AC resistance bridge, we performed a 4-lead resistance measurement of the copper trace at 300~K and 4~K, taking the ratio as the \textit{RRR} value. The length of the measured trace was chosen to ensure that the smaller 4~K resistance would remain well within a viable range for accurate measurement with the Lakeshore 372. 

\begin{table}[!h]
    \centering
    \caption{Thermal loading through the stripline between two sets of temperature stages: 300~K to 40~K, and 40~K to 4~K.\label{tab:load_estimates}}
    \begin{tabular}{lcc}
    \hline
     & Mod-Cam & Prime-Cam \\
     & Stripline & Stripline\\
     & Estimated & Predicted\\
     & Loading (W) &Loading (W)\\
    \hline
    Per Network (300~K to 40~K)& $2.0*10^{-2}$ & $7.3*10^{-3}$\\
    Per Stripline (300~K to 40~K)& $1.2*10^{-1}$ & $4.4*10^{-2}$\\
    \textbf{Per Readout Harness (300~K to 40~K)}& \textbf{0.71} & \textbf{0.79}\\
    \hline
    Per Network (40~K to 4~K)& $1.8*10^{-2}$ & $4.9*10^{-3}$\\
    Per Stripline (40~K to 4~K)& $1.1*10^{-1}$ & $3.0*10^{-2}$\\
    \textbf{Per Readout Harness (40~K to 4~K)}& \textbf{0.64} & \textbf{0.53}\\
    \end{tabular}

\end{table}

The 300~K resistance of the segment of copper trace was measured at $275~\mathrm{m}\Omega$, while the 4~K resistance was measured at approximately $2.3~\mathrm{m}\Omega$. We assume that the gold plating on the copper trace negligibly contributes to the overall trace resistance. The resulting ratio of resistances yields a \textit{RRR}~$\approx  120$, indicating the trace is made of high-purity OFHC copper \cite{nist_copper}. 

Using the measured \textit{RRR} value for these circuits and the functional form of cryogenic OFHC copper thermal conductivity given by Simon et al. \cite{nist_copper}, we obtain estimates for the thermal loading in Mod-Cam, as well as extrapolated predictions for Prime-Cam, for the configurations described in Section \ref{section:intro}. The estimates are obtained using the equation for linear thermal conductivity:
\begin{equation}
\label{eq:linear_therm_cond}
    k  = \frac{Pd}{A \Delta T},
\end{equation}
where $k$ is the thermal conductivity of the material, $P$ is the transmitted power between two ends of the material, $d$ is the distance between the ends, $A$ is the cross-sectional area of the material, and $\Delta T$ is the temperature difference between the two ends. 

For Mod-Cam, $\Delta T$ is given by real measurements of the Mod-Cam temperature stages, where the temperature of the 4~K readout mount is consistently about 12~K while the 40~K readout mount is consistently about 51~K. These elevated temperatures are known to result from poor heat sinking at various points across the 4~K  and 40~K shells. Elevated 4~K LNA temperatures have not yet impacted in-lab testing of the instrument module or detector noise characterization. 

Based on thermal budgets and loading estimates for Prime-Cam, we anticipate a realistic temperature of the 4~K readout harness will be 4~K, while the 40~K harness will be no warmer than 53~K. Initial cryogenic validation testing with Prime-Cam in the lab indicates performance will be on-par with these baseline values. 

The thermal conductivity used in these calculations $k$ assumes a linear gradient between two temperatures and is obtained by numerically integrating the functional form of $k(T)$ between the two temperatures to obtain $k_{avg}$ in the given range. For Mod-Cam estimates, $A$ is given by the profilometry measurements shown in Section \ref{section:stripline_design} and $d = 0.115~\mathrm{m}$ is set by the physical design of the stripline. For Prime-Cam predictions, the physical dimensions used reflect the changes both in the trace thickness and the distance between temperature stages outlined in Section \ref{section:intro}. 
The estimates obtained for the thermal loading from stripline in Mod-Cam, as well as predictions for Prime-Cam, are compiled in Table \ref{tab:load_estimates}.

For both instruments, the total expected thermal contributions from the stripline compiled in Table \ref{tab:load_estimates} are well within the available thermal budget at 40~K. In both instruments, the cooling power available at 4~K is more limited. The aforementioned changes to the Prime-Cam stripline are motivated by the fact that a single Prime-Cam readout harness has 3x the total number of stripline as Mod-Cam; without modification to the stripline, the predicted loading at 4~K for a single Prime-Cam readout harness would therefore be $3*0.64~W= 1.92~W$. For a fully populated Prime-Cam with 3 readout harnesses this loading at $4~K$ scales to $5-6 ~W$, far exceeding the total available cooling power. However, after redesign, the predicted loading through the stripline for a single readout harness at 4~K in Prime-Cam is $0.53 ~W$ (see Table \ref{tab:load_estimates}); for 3 harnesses this scales to approximately $1.5~W$, which is within the available thermal budget. We also note that the power dissipation from KID biasing in both the Mod-Cam and Prime-Cam stripline is expected to be at least three orders of magnitude smaller than the estimated thermal loading through the stripline. 

\subsection{Thermal Conductivity Measurement}

\begin{figure}[!h]
\includegraphics[width=\linewidth]{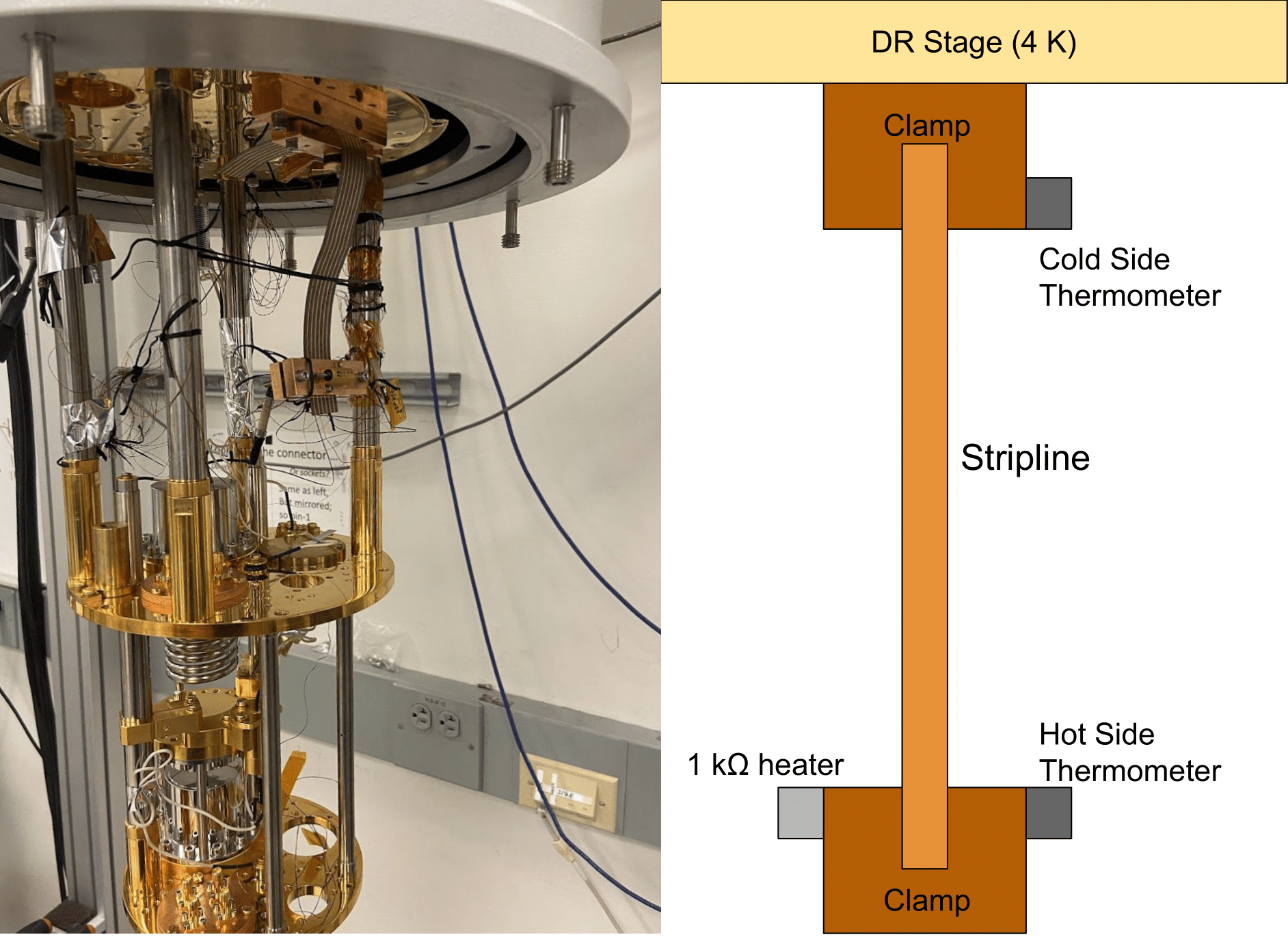}
\caption{A photograph of the testing setup used for 4~K thermal loading measurements is shown (left) along with a schematic diagram of the testing setup (right). Hot and cold side temperatures, $T_h$ and $T_c$ respectively, are each measured using a Cernox cryogenic temperature sensor. Thermal loads are applied across the stripline with a $1~\mathrm{k} \Omega$ resistor sunk to the warm side clamp.}
\label{fig:thermal_config}
\end{figure}

Thermal conductivity measurements were performed in a Bluefors SD250 Dilution Refrigerator with a custom mount allowing the stripline to be suspended from the 4~K stage. The mounting scheme is described in detail in Fig. \ref{fig:thermal_config}. To perform the measurement, we applied a current to the hot-side heater while measuring the temperatures of the hot and cold sides of the stripline, $T_h$ and $T_c$ respectively, with Cernox 1050-CD thermometers. This process was repeated for various heater powers while the stripline was heat sunk at 4~K on the cold side. Heater powers were ramped to achieve temperature gradients greater than the $\Delta T$ between stages in Mod-Cam. 

During testing, thermometers and heaters were equipped with long manganin cables to ensure the primary path for heating was through the stripline itself. Estimates of radiative power transfer to the stripline are negligible compared to the powers applied through the heater. 

As in \cite{2018_spt_readout}, we model our stripline thermal conductance, the product of the stripline thermal conductivity $k$ and cross-sectional area $A$, as a function of temperature: 
\begin{equation}
\label{eq:kA_model}
    kA = A_0 T^\alpha
\end{equation}. 

By inserting Eq. \ref{eq:kA_model} into the differential form of the 1-D heat transport equation and integrating from each end of the stripline, we obtain an expression for applied power $P$ readily fit to our measured data. 

\begin{equation}
\label{eq:fitting_eq}
P = \frac{A_0}{L(1+\alpha)}(T_h^{1+\alpha} - T_c^{1+\alpha})\\
\end{equation}
%\begin{split}
%\int Pdx &=\int A_0 T^\alpha dT \\
%PL &=\int_{Tc}^{T_h} A_0 T^\alpha dT\\

%\end{split}

\begin{figure}[!h]
\includegraphics[width=\linewidth]{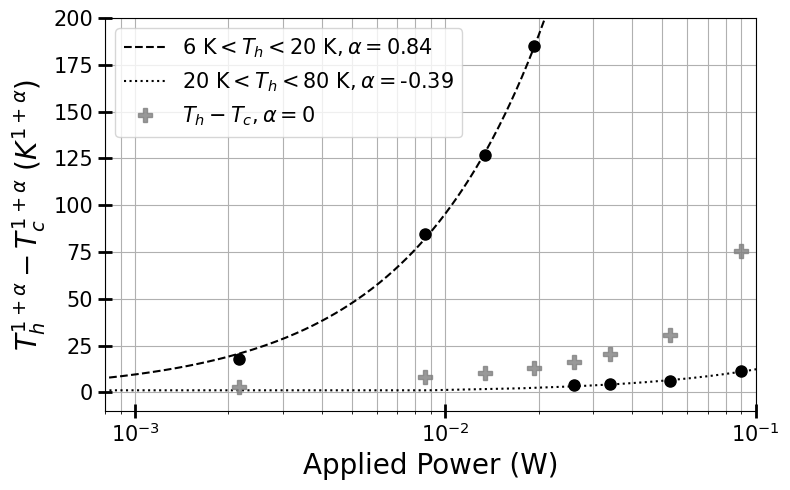}
\includegraphics[width=\linewidth]{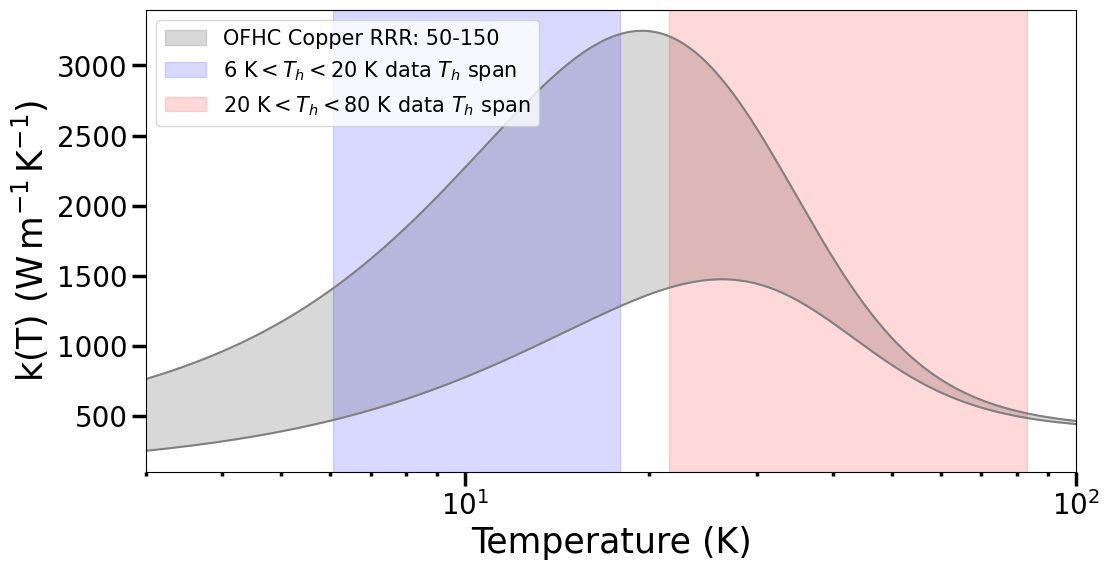}
\caption{\textit{Top}: Data points relating the power applied to the stripline to the difference in temperature between the hot and cold sides are shown with best fit lines overlaid. The data is split according to $T_h$ values occurring before and after the turnaround in OFHC copper $k(T)$ given by \cite{nist_copper}. The full dataset, containing $T_c$ and $T_h$ temperatures spanning from $3.5-84~K$ yields a best fit $kA~=~0.3\pm0.2~T^{-0.3\pm0.1}~\mathrm{mW~m~K^{-1}}$. Data points with $\alpha = 0$ corresponding to $T_h - T_c$ are shown to demonstrate the effect of rescaling. \textit{Bottom}: The ranges of $T_h$ for each half of the split dataset are overlaid on the functional form of OFHC copper thermal conductivity given by \cite{nist_copper}.}

\label{fig:thermal_results}
\end{figure}

In our data, $T_c$ spans from $3.5~\mathrm{K} - 7.5~\mathrm{K}$, while $T_h$ spans from ${6~\mathrm{K} - 84~\mathrm{K}}$. The fitted power in Eq. \ref{eq:fitting_eq} is therefore more sensitive to the larger measured changes in $T_h$. Additionally, over the range of $T_h$, the OFHC copper in the stripline traces is not well modeled as a simple power law. To compensate, we split our dataset in half and fit two regions separately for two sets of $A_0, \alpha$ parameters. The dataset is split according to the turnaround point of OFHC copper $k(T)$ at $T \approx 20~\mathrm{K}$, shown in the bottom panel of Fig. \ref{fig:thermal_results}. The low-temperature fit region contains $T_h$ from $6~\mathrm{K} - 20~\mathrm{K}$, while the higher temperature region contains $T_h$ from approximately $20~\mathrm{K} - 80~\mathrm{K}$. For the two fit regions we obtain ${kA=22\pm6T^{0.84\pm0.09}~\mathrm{\mu W~m~K^{-1}}}$ for  ${6~\mathrm{K}<T_h<20~\mathrm{K}}$ and ${kA=0.6\pm0.3T^{-0.4\pm0.1}~\mathrm{mW~m~K^{-1}}}$ for the extended ${20~\mathrm{K}<T_h<80~\mathrm{K}}$. Combined, these fits provide good characterization of the stripline thermal conductance $kA$ over the full range of hot and cold side temperatures measured, spanning from approximately ${4~\mathrm{K}-80~\mathrm{K}}$.

%Using Eq. \ref{eq:linear_therm_cond} and the temperature gradient measured across the stripline for each applied power, the thermal conductivity as a function of average temperature of the hot and cold sides of the stripline is found. Geometric factors used in Eq. \ref{eq:linear_therm_cond} are given by profilometry measurements in Section \ref{section:stripline_design} and measurements of the distance between clamps. At cryogenic temperatures, the thermal conductivity of polyimide is known to be more than 4000x smaller than that of copper \cite{kapton_thermal}, so thermal conduction through the polyimide dielectric is assumed to be negligible compared to the circuit copper layers. The result of the measurement is shown in Fig \ref{fig:thermal_results}. The direct measurements of thermal conductivity agree with calculated values using standard $RRR$ values for OFHC copper in line with results reported in Section \ref{section:RRR_result}.

\section{Crosstalk}\label{section:crosstalk}

\subsection{Measurement}
During in-lab testing of the 280~GHz instrument module we observed elevated crosstalk through full readout chains. Cold measurements confirmed no significant reflections in the readout chains, so we are confident excess pickup is primarily occurring between different readout chains. We sometimes refer to this as inter-channel crosstalk. In order to isolate the source of this crosstalk, we performed warm S21 measurements using an Agilent E5071C VNA over the full readout bandwidth for the TiN and Al arrays installed in the 280~GHz module, which combined span a range from 300~MHz - 1.5~GHz. Crosstalk measurements were performed for a variety of configurations: a lone stripline, a stripline with a SMP-SMA transition PCB connected on either end (mimicking the readout structure in Mod-Cam shown in Fig. \ref{fig:pcb_picture}), and a lone transition PCB.

\subsection{Crosstalk Measurement Results}

\begin{figure}[!h]
\includegraphics[width=\linewidth]{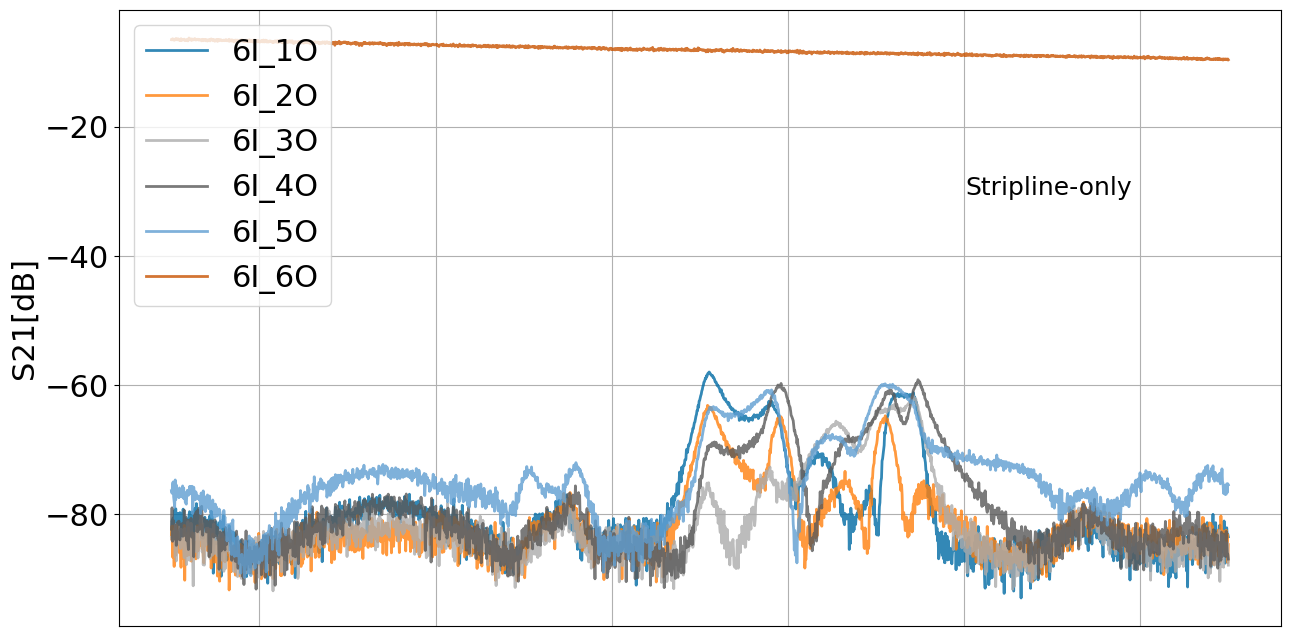}
\includegraphics[width=\linewidth]{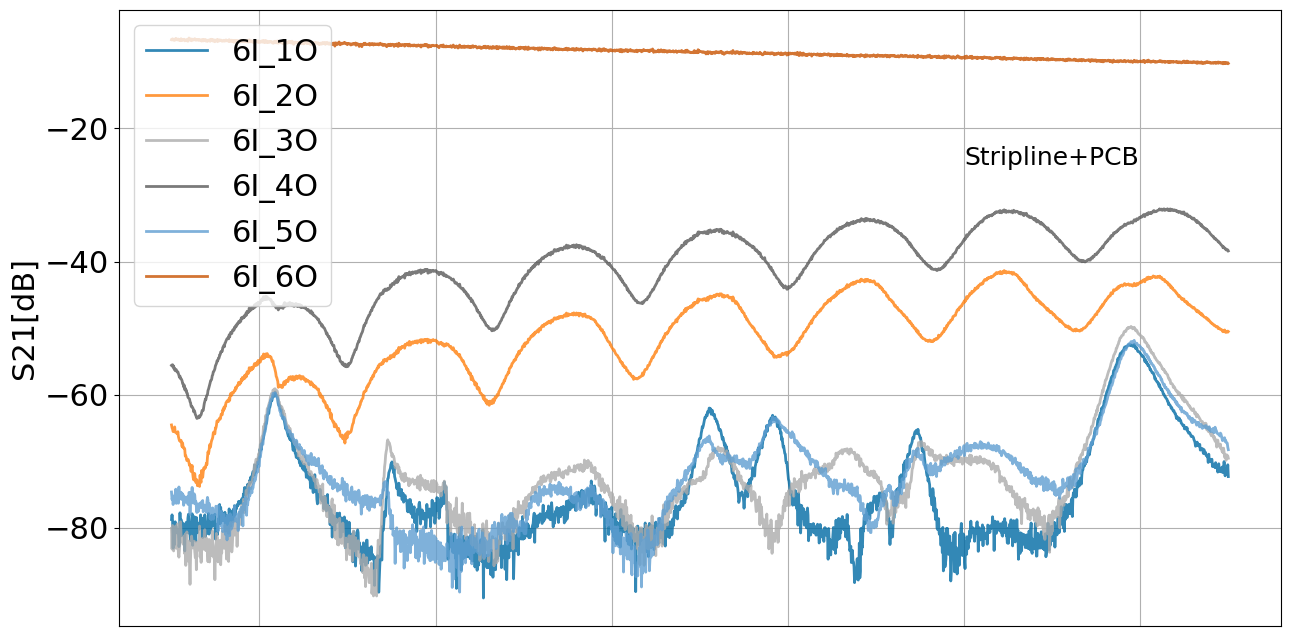}
\includegraphics[width=\linewidth]{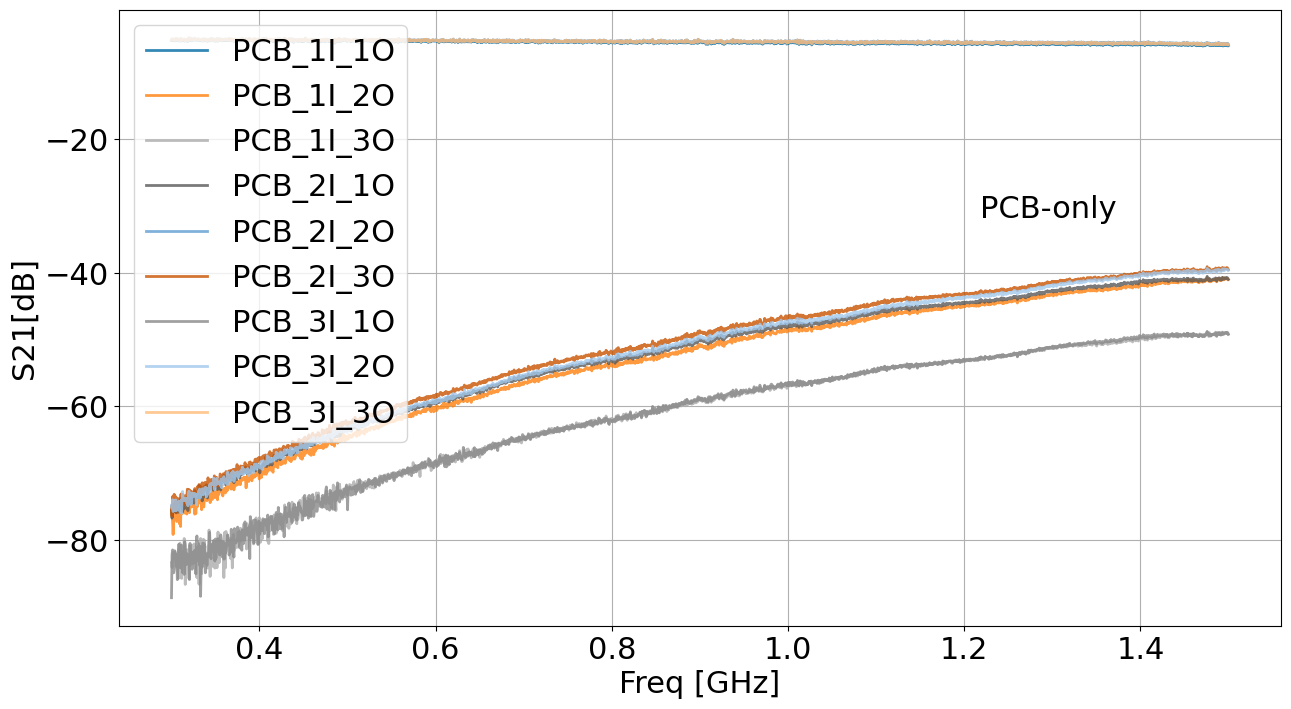}
\caption{
Warm crosstalk measurements are shown. \textit{Top}: Crosstalk measured for all channel combinations with a single channel (6) on an isolated stripline. The labeling scheme, following that shown in Fig. \ref{fig:pcb_picture}, indicates the Input (I) and Output (O) with each associated channel number. All inter-channel measurements show comparable S21 amplitude, while transmission through channel 6 is observed at the top of the panel. \textit{Middle}: Crosstalk measured through the combination of a stripline with SMP-SMA transition PCBs on either end. The two elevated traces with uniform standing wave patterns believed to result from connectors correspond to traces that share a connector, and hence also an SMP-SMA transition PCB. \textit{Bottom}: Crosstalk measured through an isolated transition PCB following the PCB labeling scheme in Fig. \ref{fig:pcb_picture}. Channels spaced physically further on the PCB show lower S21 amplitude than those immediately adjacent. 
}

\label{fig:crosstalk}
\end{figure}

Warm S21 measurements of the configurations are shown in Fig. \ref{fig:crosstalk}, with labels corresponding to connections designated in the labeling scheme in Fig. \ref{fig:pcb_picture}. The results demonstrate that the stripline alone is not the dominant source of crosstalk in the readout chain, with inter-channel crosstalk shown in the top panel of Fig. \ref{fig:crosstalk} uniformly 50~dB or more below transmission, meeting project requirements. Additional structure visible in the inter-channel traces on this plot are primarily attributed to the SMP connections between the stripline and PCBs; this structure was observed to shift slightly when connectors were manually perturbed.

The middle ``stripline~+~PCB" panel of Fig. \ref{fig:crosstalk} isolates the source of crosstalk in the readout chain to the custom transition PCBs; in this configuration, chains 2, 4, and 6 share the same PCB and show elevated crosstalk, up to 30~dB above those combinations that do not share a PCB. The level of crosstalk is also correlated with physical spacing on the transition PCB, accounting for the elevated level of the $\mathrm{6I\_4O}$  measurement relative to the $\mathrm{6I\_2O}$ measurement.

Inter-channel crosstalk measurements in the ``PCB-only'' panel of Fig. \ref{fig:crosstalk} also confirm the SMP-SMA transition PCB as the source of elevated crosstalk. The crosstalk in trace $PCB\_\mathrm{3I\_2O}$ for a single PCB is  $-40~dB$ at 1.4 GHz. For the full chain two PCBs are used, resulting in the $3~ dB$ higher baseline crosstalk demonstrated in the ``stripline~+~PCB" panel of Fig. \ref{fig:crosstalk} in the corresponding trace labeled $\mathrm{6I\_2O}$. In light of this result, the SMP-SMA transition PCBs have been redesigned using an impedance matched via to quickly bury the signal trace in an intermediate PCB layer between ground planes for improved isolation. Radiative effects from the via ends are expected to be minimal from signals in the readout bandwidth. %Moreover, our ability to isolate signal traces on the PCB is fundamentally limited by the ganged SMP connector design, which is necessarily mounted on the edge of the PCB board. At minimum, this results in a small section of microstrip trace on the PCB which can add crosstalk. The original PCB design referenced in Fig. \ref{fig:pcb_picture} and Fig. \ref{fig:crosstalk} utilized only microstrip traces. 

To estimate the predicted crosstalk of the new PCB design we performed simulations of the via-trace structure on the PCB using Ansys HFSS software, with critical dimensions selected to impedance match the full structure to  $50 ~\Omega$. These simulations estimate crosstalk at a level no greater than -60~dB for a single structure, and -57~dB across the full PCB. This represents an approximately 20~dB improvement in crosstalk level over the prior PCB design, allowing the PCB to exceed project requirements. 

\section{Conclusion}
We performed two types of measurements to characterize the thermal loading through the stripline flexible circuitry designed for Mod-Cam and Prime-Cam: finding the $RRR$ of the copper traces and directly estimating the thermal conductivity of the circuit as a function of temperature for various temperature ranges. Using the $RRR$ result, we predicted the expected thermal loading for nominal stripline configurations for both Mod-Cam and Prime-Cam. Results of these predictions demonstrate the significant improvement in expected loading for the redesigned Prime-Cam stripline circuits.

When in-lab testing of the readout in Mod-Cam yielded higher-than-anticipated crosstalk through the readout chains, we performed measurements to isolate the source of this pickup. Results of this testing identified the source as the SMP-SMA transition PCBs, which carry signals from the stripline flexible circuits into coaxial cables. Subsequent redesign and simulation of these PCBs has demonstrated that they should meet project requirements for low crosstalk when fabricated. This new PCB design should be implemented in both Mod-Cam and Prime-Cam.  

Flexible stripline circuits installed in Mod-Cam have proven to be a robust option for the simultaneous readout of thousands of KIDs during months of in-lab testing. The results presented in this paper validate stripline as a scalable solution for readout of large-format KID arrays and inform the ongoing fabrication of optimized Prime-Cam flexible circuit assemblies. 

\section*{Acknowledgments}
The CCAT project, FYST and Prime-Cam instrument have been supported by generous contributions from the Fred M. Young, Jr. Charitable Trust, Cornell University, and the Canada Foundation for Innovation and the Provinces of Ontario, Alberta, and British Columbia. The construction of EoR-Spec is supported by NSF grant AST-2009767. The construction of the 350 GHz instrument module for Prime-Cam is supported by NSF grant AST-2117631. The construction of the FYST telescope was supported by the Gro{\ss}ger{\"a}te-Programm of the German Science Foundation (Deutsche Forschungsgemeinschaft, DFG) under grant INST 216/733-1 FUGG, as well as funding from Universit{\"a}t zu K{\"o}ln, Universit{\"a}t Bonn and the Max Planck Institut f{\"u}r Astrophysik, Garching.

S. Walker acknowledges support from the National Science Foundation under Award No. 2503181.

\ifCLASSOPTIONcaptionsoff
  \newpage
\fi

\bibliographystyle{IEEEtran}

\nocite{*}

\bibliography{bibtex/bib/bib}

\end{document}